# A Planarity Test via Construction Sequences

Jens M. Schmidt

MPI für Informatik, Saarbrücken, Germany

**Abstract**

Optimal linear-time algorithms for testing the planarity of a graph are well-known for over 35 years. However, these algorithms are quite involved and recent publications still try to give simpler linear-time tests. We give a simple reduction from planarity testing to the problem of computing a certain construction of a 3-connected graph. The approach is different from previous planarity tests; as key concept, we maintain a planar embedding that is 3-connected at each point in time. The algorithm runs in linear time and computes a planar embedding if the input graph is planar and a Kuratowski-subdivision otherwise.

## 1 Introduction

Testing the planarity of a graph is a fundamental algorithmic problem that has initiated significant contributions to data structures and the design of algorithms in the past. Although optimal linear-time algorithms for this problem are known for over 35 years [12, 3], they are involved and recent publications still try to give simpler linear-time algorithms [4, 5, 8, 10, 23].

The aim of this paper is a linear-time planarity test that is based on a simple reduction to the problem of computing a certain construction $C$ of a 3-connected graph (we will give a precise definition of $C$ in Chapter 3). The existence of the construction $C$ has also been used by Kelmans [13] and Thomassen [25] to give a short proof of Kuratowski's Theorem. Although their proof itself is constructive (in the sense that it gives a polynomial-time planarity test) and received much attention in graph theory due to its simplicity, it has not been utilized algorithmically to establish a linear-time planarity test. Our hope is that the linear-time algorithm presented here will lead to simple planarity tests just as the same concept led to simple proofs of Kuratowski's Theorem.

Currently, the fastest algorithm known for computing $C$ achieves a linear running time [21], but is quite involved. For that reason, the resulting planarity test does not qualify to be regarded as simple yet. However, every simplification made for computing $C$ will immediately result in a simpler linear-time planarity test. If we allow a quadratic running time, a very simple algorithm that computes $C$ (and thus planarity) is known [22].

Recent planarity tests like [4, 5, 8, 10, 23] maintain a planar embedding at each step, where all steps either add paths/edges (*path addition method*) or vertices (*vertex addition method*) to the embedding (for an extensive survey, we refer to the chapter of Patrignani in [24]). In our algorithm, each step will essentially add an edge, whose endpoints might subdivide other edges before.

Unlike previous planarity tests, we maintain a planar embedding that is always 3-connected. This is a key concept for the following reason. The 3-connectivity constraint fixes the planar embedding (up to flipping), which will allow to test very easily whether the addition of a next edge $e$ preserves planarity.

A planarity test can be made *certifying* in the sense of [15] by augmenting its *yes/no*-output with a planar embedding if the input graph is planar and a Kuratowski-subdivision otherwise. The first two linear-time planarity tests of Hopcroft and Tarjan [12] and Booth and Lueker [3] did not give a planar embedding for planar input graphs, respectively. Mehlhorn and Mutzel [17] and Chiba, Nishizeki, Abe and Ozawa [6], respectively, extended these tests to compute a planar embedding in the same asymptotic running time. The algorithm presented here is certifying.

## 2 Preliminaries

We use standard graph-theoretic terminology from [2]. Let $G = (V, E)$ be a simple finite graph with $n := |V|$ and $m := |E|$. Multiedges do not matter for planarity and can be removed in advance by performing two bucket sorts on the endpoints of edges in $E$.

A vertex whose deletion increases the number of connected components is called a *cut vertex*. A graph $G$ is *biconnected* if it is connected and contains no cut vertex. A *biconnected component* of a graph $G$ is a maximal biconnected subgraph of $G$. A pair of vertices whose deletion disconnects a graph is called a *separation pair*. A biconnected graph is *triconnected* if it contains no separation pair. A *subdivision* of a graph $G$ (a $G$-*subdivision*) is a graph obtained by replacing the edges of $G$ with internally disjoint paths of length at least one. Triconnected graphs and their subdivisions have the following property, which we will use throughout this paper.

**Lemma 1** (Whitney [29], Theorem 1.1 in [19])**.** *Every subdivision of a triconnected graph has a unique planar embedding (up to flipping).*

The *triconnected components* of a graph $G$ are obtained by the following recursive process on every biconnected component $H$ of $G$: As long as there is a separation pair $\{x, y\}$ in $H$, we split $H$ into two subgraphs $H_1$ and $H_2$ that partition $E(H)$ and have only $x$ and $y$ in common, followed by adding the edge $e = xy$ to both $H_1$ and $H_2$. We refer to [9] for a precise definition of this process. The edge $e$ that was added to $H_1$ (respectively, $H_2$) is called the *virtual edge* of $H_1$ ($H_2$) and can be seen as a replacement of the graph $H_2$ ($H_1$) in this decomposition.

The graphs resulting from this process are either sets of three parallel edges (*triple-bonds*), triangles or simple triconnected graphs. To obtain the triconnected components of $G$, triple-bonds containing the same virtual edge are successively merged to maximal sets of parallel edges (*bonds*); similarly, triangles containing the same virtual edge are successively merged to maximal cycles (*polygons*). Thus, a triconnected component of $G$ is either a bond, a polygon, or a simple triconnected graph.

It is well-known that a graph $G$ is planar if and only if all its biconnected components are planar [12]. A similar result holds for the triconnected components of $G$: If $G$ is planar, all triconnected components of $G$ are planar, as every triconnected component is a minor of $G$. Conversely, if all triconnected components of a graph $G$ are planar, we can successively merge the planar embeddings of two triconnected components containing the same virtual edge to a bigger planar embedding [28, Lemma 6.2.6], and obtain a planar embedding for $G$ in linear time. This gives the following result.

**Lemma 2** ([16])**.** *A graph is planar if and only if all its triconnected components are planar.*

As bonds and cycles are planar, planarity has only to be checked for simple triconnected graphs. The triconnected components can be computed in linear time [11, 9] and reliable implementations are publicly available [20].

# 3 Constructions of Triconnected Graphs

With the above arguments we can assume that the input graph $G$ is simple and triconnected. We will make use of a special construction of triconnected graphs due to Barnette and Grünbaum [1].

**Definition 3.** Let $G$ be a simple triconnected graph with $n \geq 4$. We define the following operations on $G$ (all vertices and edges are assumed to be distinct; see Figure 1).

(a) Add an edge $xy$ between two non-adjacent vertices $x$ and $y$.

(b) Subdivide an edge $ab$ by a vertex $x$ and add the edge $xy$ for a vertex $y \notin \{a, b\}$.

(c) Subdivide two non-parallel edges $e$ and $f$ by vertices $x$ and $y$, respectively, and add the edge $xy$ (note that $e$ and $f$ may intersect in one vertex).

(d) Add a new vertex $x$ and join it to exactly three old vertices $a$, $b$ and $c$.

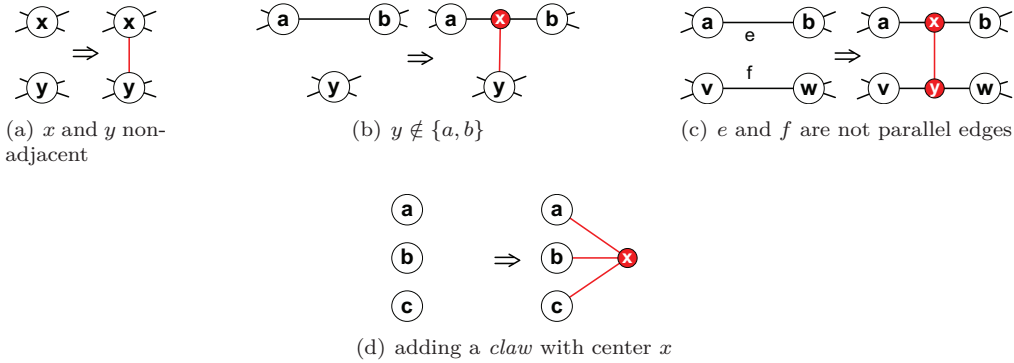

(a) $x$ and $y$ non-adjacent

(b) $y \notin \{a, b\}$

(c) $e$ and $f$ are not parallel edges

(d) adding a *claw* with center $x$

Figure 1: Operations on triconnected graphs

Operations 3a-c correspond to *adding edges* (the *added* edge is $xy$) while Operation 3d corresponds to *adding a claw* (i.e., $K_{1,3}$) with a designated *center vertex* $x$. The *attachments* of an operation $O$ on $G$ are the vertices and edges in $G$ involved in the operation, i.e., the attachments of Operations 3a–d are $\{x, y\}$, $\{ab, y\}$, $\{ab, vw\}$ and $\{a, b, c\}$, respectively. Let *suppressing* a vertex $x$ with exactly two neighbors $y$ and $z$ be the operation of deleting $x$ and adding the edge $yz$.

Applying any of the Operations 3a–d to $G$ generates a graph that is simple and triconnected again. A classical result of Barnette and Grünbaum [1] and Tutte [26] characterizes the triconnected graphs in terms of the first three operations.

**Theorem 4** ([1, 26])**.** *A simple graph $G$ with $n \geq 4$ is triconnected if and only if $G$ can be constructed from $K_4$ using only operations of Types 3a–c.*

For testing planarity, we will use the following slightly modified construction. It restricts operations of Type 3a to be at the end of the construction and uses additional operations of Type 3d. This will allow for an easier efficient data structure in the planarity test.

**Theorem 5** ([22])**.** *A simple graph $G$ with $n \geq 4$ is triconnected if and only if $G$ can be constructed from $K_4$ using only operations of Types 3a–d such that all operations of Type 3a are applied last.*

A *construction sequence* $C$ of $G$ is a sequence of operations that constructs $G$ from $K_4$ precisely as stated in Theorem 5. Note that any edge that was added by an Operation 3a in $C$ will not be subdivided by later operations.

A construction sequence can be represented in space linearly dependent on the size of $G$ by using a labeling scheme on vertices and edges that essentially assigns a new label to one half of an edge $e$ after $e$ was subdivided by an operation [22]. The labeling scheme allows additionally for a constant-time access to the edges and vertices that are involved in an operation $O$, i.e., to the edge $e$ that is added by $O$ and to the vertices and edges on which the endpoints of $e$ lie.

Recently, it was shown that a construction $C'$ of $G$ as stated in Theorem 4 can be computed in linear time [21]. Although the algorithm is somewhat involved, it is certifying and an implementation has already been made publicly available [18]. A construction sequence $C$ can be obtained from $C'$ by a simple linear-time transformation as pointed out in [22].

## 4 The Planarity Test

We can assume that the input graph $G$ is simple and triconnected. If $n \leq 3$, $G$ is planar, so assume $n \geq 4$. Let $C$ be a construction sequence of $G$.

The planarity test starts with the (unique) planar embedding of $K_4$ and computes iteratively a planar embedding for the graph that is obtained from the next operation $O$ in $C$ if possible. We have to know under which conditions an operation $O$ in $C$ preserves planarity.

**Lemma 6.** *Let $H$ be a planar embedding of a simple triconnected graph on at least 4 vertices and let $H'$ be the graph that is obtained from $H$ by applying an operation $O$ of Type 3a–d. Then $H'$ is planar if and only if all attachments of $O$ are contained in one face $f$ of $H$.*

*Proof.* $\Rightarrow$: Let $H'$ be planar and assume to the contrary that $O$ has two attachments $a$ and $b$ that are not contained in one face of $H$. In particular, $H'$ has a planar embedding $Emb'$. Let $Emb$ be the planar embedding that is obtained from $Emb'$ by inversing Operation $O$, i.e., by deleting the added edge in $Emb'$ and suppressing all vertices of degree two if $O$ is of Type 3a–c, and by deleting the center vertex of the added claw in $Emb'$ if $O$ is of Type 3d. Then $Emb$ and $H$ embed the same simple triconnected graph. Moreover, $Emb$ and $H$ are combinatorially different embeddings, as $Emb$ has a face containing both $a$ and $b$, while $H$ has no face containing both $a$ and $b$. This contradicts Lemma 1.

$\Leftarrow$: Clearly, subdividing edges in the face cycle of $f$ preserves planarity and so does the addition of an edge inside $f$ that has the new vertices of the subdivided edges as endpoints. Similarly, adding a claw inside $f$ preserves planarity when all its attachments are contained in $f$. Thus, applying $O$ results in a planar embedding for $H'$, which proves that $H'$ is planar. $\square$

Clearly, if all the attachments of $O$ are in $f$, adding the new edge or vertex of $O$ inside $f$ gives immediately a planar embedding of $G$. If all operations in $C$ preserve planarity, we obtain a planar embedding of $G$. Otherwise, $G$ contains a Kuratowski-subdivision; we will show how to extract this subdivision in linear time in the next chapter. Lemma 6 suggests the following Algorithm 1.

It remains to discuss how the condition in Lemma 6 can be checked efficiently for every operation in $C$.

A *plane st-graph* is an embedding of a directed acyclic graph with exactly one source $s$ and exactly one sink $t$ such that $s$ and $t$ are contained in the external face of the

---
**Algorithm 1** PlanarityTest(G)  ▷ $G$ simple and triconnected with $n \geq 4$
---
1: compute a construction sequence $C = O_1, \ldots, O_k$ of $G$
2: initialize the (unique) planar embedding $H$ of $K_4$
3: **for** $i = 1$ **to** $k$ **do**
4:     **if** all attachments of $O_i$ are in one face $f$ of $H$ **then** ▷ planar
5:         apply $O_i$ to $H$ by adding the edge or claw inside $f$
6:     **else** ▷ non-planar
7:         compute a Kuratowski-subdivision
---

embedding. It is well-known that every biconnected graph can be drawn as plane $st$-graph. The plane embedding $H$ is triconnected and thus biconnected in every step of algorithm 1. To check the condition in Lemma 6 efficiently, we will maintain $H$ as plane $st$-graph and use a data structure that is able to answers queries whether edges and vertices are contained in the same face of $H$ in amortized constant time.

The data structure is due to Djidjev [7] and runs on a standard word-RAM. It maintains a plane $st$-graph $H$ in which the incoming and the outgoing edges for any vertex $x$ appear consecutively around $x$; hence, the boundary of each face $f$ in $H$ consists of two oriented paths from a common start vertex (the *source* of $f$) to a common end vertex (the *sink* of $f$). Additionally, every vertex $x \notin \{s,t\}$ is contained in exactly two faces for which $x$ is neither source nor sink; we call these faces the *left* and the *right* face of $x$, respectively. The data structure maintains pointers to the source and sink for each face in $H$ and a pointer from each vertex $x \notin \{s,t\}$ to its left and right face. The following queries for triconnected graphs $H$ are supported.

(1) Given a vertex $a$ and an edge $b$ of $H$, output a face of $H$ that contains $a$ and $b$ or report that there is none.

(2) Given two vertices $a$ and $b$ of $H$ such that $a$ is source or sink in at most 11 faces, output a face of $H$ that contains $a$ and $b$ or report that there is none.

Each of the queries takes worst-case time $O(1)$. We show that also the following query can be computed in worst-case time $O(1)$.

(3) Given three vertices $a$, $b$ and $c$ of $H$, output a face of $H$ that contains $a$, $b$ and $c$ or report that there is none.

We can compute the set $F$ of all left and right faces of the vertices $a$, $b$ and $c$ in constant time. If there is a face $f$ in $H$ containing $a$, $b$ and $c$, at least one vertex in $\{a,b,c\}$ is neither source nor sink of $f$, which implies that $f \in F$. It therefore suffices for a query (3) to test each face $f \in F$ on containing $a$, $b$ and $c$, respectively. A vertex $v$ is contained in $f$ if and only if $v$ is either source or sink of $f$, which can be checked in time $O(1)$, or one of the remaining vertices in $f$, which can be checked in time $O(1)$ by testing whether $f$ is left or right face of $v$.

The data structure supports additionally each of the following modifications to $H$ in amortized time $O(1)$ and maintains a plane $st$-graph after every modification.

(4) Subdivide an edge.

(5) Given two non-adjacent vertices $a$ and $b$ and a face $f$ of $H$ that contains $a$ and $b$, add the edge $ab$ inside $f$.

Clearly, $K_4$ can be embedded as a plane $st$-graph and we initialize $H$ with this embedding. Every operation $O$ of Type 3a–d can be converted into at most three of the modifications (4) and (5). E. g., we can add a claw having its attachments $\{a, b, c\}$ in a common face by consecutively inserting the edge $ab$, subdividing $ab$ with a new vertex $x$ and adding the edge $xc$. For operations $O$ of Type 3b–d, the condition in Lemma 6 can be checked in constant time by one query (1) or one query (3).

It only remains to show how we can check the condition in Lemma 6 if $O$ is of Type 3a. According to Theorem 5, all operations in $C$ that follow $O$ will be of Type 3a, which implies that $H$ is a spanning subgraph of $G$. In other words, each of the remaining operations in $C$ adds only an edge that will not be subdivided. Hence, the order in which these remaining edges $E'$ are added does not matter; we sort them lexicographically according to their endpoints.

We use a trick given by Djidjev (Lemma 3.2 in [7]) and maintain an auxiliary graph $H^A$ whose vertex set consists of all vertices in $V(H)$ that are source or sink in some face of $H$ and which are incident to an edge in $E'$. There is an edge between two vertices $a$ and $b$ in $H^A$ if $a$ and $b$ are source and sink vertices of the same face. The graph $H^A$ can be constructed in linear time when the first operation 3a in $C$ is encountered; after every modification (5), $H^A$ can be updated in time $O(1)$, as each face $f$ stores a pointer to its source and sink.

As $H^A$ is planar and has at most two parallel edges between every two vertices (as $H$ is simple and triconnected), it contains at most $6|V(H^A)| - 12$ edges. Hence, there is at least one vertex of degree at most 11 in $H^A$. Before the first Operation 3a in $C$ is applied to $H$, we construct a list $Small$ of all vertices in $H^A$ having degree at most 5 in linear time; again, this list is easy to maintain under modifications (5) in time $O(1)$. Now we just choose successively a vertex $v \in Small$ and an edge $e = vw$ in $E'$ and perform modification (5) with $v$ and $w$ (if $v$ and $w$ have been reported to be in the same face). This allows to check the condition in Lemma 6 for each of the remaining edges in $E'$ in constant time with query (2). We conclude the following theorem.

**Theorem 7.** *The planarity test Algorithm 1 can be implemented in linear time.*

## 5 Extensions

We show how a Kuratowski-subdivision can be computed if an operation $O$ is encountered that has not all attachment vertices on one face in $H$. For the computation, we will go along the arguments given in the short proof of Kuratowski's Theorem in [25].

We first recall planarity-related terminology. For a cycle $C$ in a graph $G$, let a *C-component* be either an edge $e \notin C$ with both endpoints in $C$ or a connected component of $G \setminus V(C)$ together with all edges that join the component to $C$ and all endpoints of these edges. The *vertices of attachment* of a $C$-component $H$ are the vertices in $H \cap C$. Two $C$-components $H_1$ and $H_2$ *avoid each other* if $C$ contains two vertices $u$ and $v$ such that $H_1$ has all vertices of attachment on one path in $C$ from $u$ to $v$ and $H_2$ has all vertices of attachment on the other path in $C$ from $u$ to $v$.

Two $C$-components *overlap* if they do not avoid each other. Let two $C$-components $H_1$ and $H_2$ be *C-equivalent* if $H_1 \cap C = H_2 \cap C$ and this set contains exactly three vertices. Let $H_1$ and $H_2$ be *skew* if $C$ contains four distinct vertices $x_1$, $x_2$, $x_3$ and $x_4$ in cyclic order such that $x_1$ and $x_3$ are in $H_1$ and $x_2$ and $x_4$ are in $H_2$. We will need the following basic fact about $C$-components.

**Lemma 8** ([25]). *Two C-components overlap if and only if they are either skew or C-equivalent.*

Now we are prepared to compute a Kuratowski-subdivision.

**Lemma 9.** *Let $H$ be a planar embedding of a simple triconnected graph on at least 4 vertices and let $O$ be an operation of Type 3a–d on $H$ whose attachments are not all* *contained in one face of $H$. Then the graph $H'$ that is obtained from $H$ by applying $O$ contains a subdivision of $K_5$ or $K_{3,3}$, which can be computed in linear time.*

*Proof.* The proof follows the arguments given in [25] and [27]. Let $a$ and $b$ be two attachments of $O$ that are not contained in one face of $H$; note that $a$ may be an edge. As $H \setminus a$ is 2-connected, it contains a cycle $C$ that is the boundary of the face which contains $a$ in its interior. By assumption, $b \notin C$. Let $H_a$ and $H_b$ be the $C$-components of $H$ containing $a$ and $b$, respectively. By definition of $C$, $H_a$ is the only $C$-component in the interior of $C$.

We show that $H_a$ and $H_b$ overlap. Assume the contrary. Then $H_b$ has two vertices of attachment $u$ and $v$ such that $H_a$ has all vertices of attachment on one path $P_a \subset C$ from $u$ to $v$ and $H_b$ has all vertices of attachment on the other path $P_b \subset C$ from $u$ to $v$. If $a$ is a vertex, $a$ and $b$ are in different components of $H \setminus \{u, v\}$, since $H$ is a planar embedding. This contradicts $H$ to be triconnected. Otherwise, $a$ is an edge (which will be subdivided by $O$) and $H_a = a$. Then, as $H$ is simple, $P_a$ has length at least two, which implies that an inner vertex in $P_a$ is in a different component of $H \setminus \{u, v\}$ than $b$. This contradicts $H$ to be triconnected. Thus, $H_a$ and $H_b$ overlap.

According to Lemma 8, $H_a$ and $H_b$ are either skew or $C$-equivalent. The cycle $C$, $H_a$ and $H_b$ can be easily computed in linear time. Deciding whether $H_a$ and $H_b$ are skew and computing the vertices $x_1$, $x_2$, $x_3$ and $x_4$ on $C$, whose existence defines this property amounts to one traversal along $C$. If $a$ is an edge, subdivide $a$ and let $a'$ be the new vertex of degree two; otherwise let $a' = a$. Define $b'$ accordingly. Due to Menger's Theorem, there are either two or three internally disjoint paths from $a'$ to $C$ in $H_a$ (and from $b'$ to $C$ in $H_b$), depending on whether $H_a$ ($H_b$) is an edge. These paths can be computed by any graph traversal such as depth-first search that starts with the desired vertex.

If $H_a$ and $H_b$ are skew, we compute two of these paths in $H_a$ that end at $x_1$ and $x_3$, respectively, and two in $H_b$ that end at $x_2$ and $x_4$, respectively. Taking the union of these paths, $C$ and $T$, where $T$ is either the added edge of $O$ or the path of length two from $a$ to $b$ if $O$ adds a claw forms a $K_{3,3}$-subdivision. If $H_a$ and $H_b$ are $C$-equivalent, the union of the three paths in $H_a$ and $H_b$, respectively, $C$ and $T$, gives a $K_5$-subdivision. □

We remark that Lemma 9 can be easily extended to output always a $K_{3,3}$-subdivision when $H' \neq K_5$ is non-planar in linear time. This is based on the following variant of Kuratowski's Theorem for triconnected graphs.

**Lemma 10** ([14]). *A simple triconnected graph $G \neq K_5$ is planar if and only if $G$ does not contain a $K_{3,3}$-subdivision.*

Note that we get a $K_5$-subdivision $K$ only in the case that $O$ adds a claw. The desired $K_{3,3}$-subdivision can then be obtained from $K$ by rerouting one of the paths of $K$ that ends at $a$ to the center vertex of the claw.

**Open Questions.** Probably the most immediate question is whether there is a simple linear-time algorithm that computes the construction sequence $C$ of a triconnected graph. This would immediately imply a simple planarity test. Further, it seems possible that such an algorithm, or the existing one in [21], can be extended to compute the triconnected components of the input graph, similarly as in the triconnectivity test of Hopcroft and Tarjan [11]. This would subsume the computation of $C$ and the preprocessing of the graph into triconnected components.